\documentclass[11pt]{article}
\usepackage{graphicx}
\usepackage{amssymb,amsmath,amsfonts}

\def\Re{{\cal R \mskip-4mu \lower.1ex \hbox{\it e}\,}}
\def\Im{{\cal I \mskip-5mu \lower.1ex \hbox{\it m}\,}}
\def\ie{{\it i.e.}}
\def\eg{{\it e.g.}}

\def\sub#1{_{\lower.25ex\hbox{$\scriptstyle#1$}}}
\def\tev{\,{\ifmmode\mathrm {TeV}\else TeV\fi}}
\def\gev{\,{\ifmmode\mathrm {GeV}\else GeV\fi}}
\def\mev{\,{\ifmmode\mathrm {MeV}\else MeV\fi}}
\def\mpl{\ifmmode M_{pl}\else $M_{pl}$\fi}
\def\mpl{\ifmmode \overline M_{Pl}\else $\bar M_{Pl}$\fi}
\def\to{\rightarrow}

\def\subw{_{\rm w}}
\def\mh{\ifmmode m\sbl H \else $m\sbl H$\fi}
\def\mch{\ifmmode m_{H^\pm} \else $m_{H^\pm}$\fi}
\def\mt{\ifmmode m_t\else $m_t$\fi}
\def\mc{\ifmmode m_c\else $m_c$\fi}
\def\mz{\ifmmode M_Z\else $M_Z$\fi}
\def\mw{\ifmmode M_W\else $M_W$\fi}
\def\mws{\ifmmode M_W^2 \else $M_W^2$\fi}
\def\mhs{\ifmmode m_H^2 \else $m_H^2$\fi}   
\def\mzs{\ifmmode M_Z^2 \else $M_Z^2$\fi}
\def\mts{\ifmmode m_t^2 \else $m_t^2$\fi}
\def\mcs{\ifmmode m_c^2 \else $m_c^2$\fi}
\def\mchs{\ifmmode m_{H^\pm}^2 \else $m_{H^\pm}^2$\fi}
\def\ztwo{\ifmmode Z_2\else $Z_2$\fi}
\def\zone{\ifmmode Z_1\else $Z_1$\fi}
\def\mtwo{\ifmmode M_2\else $M_2$\fi}
\def\mone{\ifmmode M_1\else $M_1$\fi}
\def\tb{\ifmmode \tan\beta \else $\tan\beta$\fi}
\def\xw{\ifmmode x\subw\else $x\subw$\fi}
\def\ch{\ifmmode H^\pm \else $H^\pm$\fi}
\def\lum{\ifmmode {\cal L}\else ${\cal L}$\fi}
\def\inpb{\,{\ifmmode {\mathrm {pb}}^{-1}\else ${\mathrm {pb}}^{-1}$\fi}}
\def\infb{\,{\ifmmode {\mathrm {fb}}^{-1}\else ${\mathrm {fb}}^{-1}$\fi}}
\def\epem{\ifmmode e^+e^-\else $e^+e^-$\fi}
\def\ppb{\ifmmode \bar pp\else $\bar pp$\fi}
\def\bsg{\ifmmode B\to X_s\gamma\else $B\to X_s\gamma$\fi}
\def\bsll{\ifmmode B\to X_s\ell^+\ell^-\else $B\to X_s\ell^+\ell^-$\fi}
\def\bstt{\ifmmode B\to X_s\tau^+\tau^-\else $B\to X_s\tau^+\tau^-$\fi}
\def\lamt{\ifmmode \tilde\lambda\else $\tilde\lambda$\fi}
\def\shat{\ifmmode \hat s\else $\hat s$\fi}
\def\that{\ifmmode \hat t\else $\hat t$\fi}
\def\uhat{\ifmmode \hat u\else $\hat u$\fi}

\newskip\zatskip \zatskip=0pt plus0pt minus0pt
\def\matth{\mathsurround=0pt}
\def\lsim{\mathrel{\mathpalette\atversim<}}
\def\gsim{\mathrel{\mathpalette\atversim>}}
\def\atversim#1#2{\lower0.7ex\vbox{\baselineskip\zatskip\lineskip\zatskip
  \lineskiplimit 0pt\ialign{$\matth#1\hfil##\hfil$\crcr#2\crcr\sim\crcr}}}

\def\grtsim{\,\,\rlap{\raise 3pt\hbox{$>$}}{\lower 3pt\hbox{$\sim$}}\,\,}
\def\lsim{\,\,\rlap{\raise 3pt\hbox{$<$}}{\lower 3pt\hbox{$\sim$}}\,\,}

\renewcommand{\thefootnote}{\fnsymbol{footnote}}

\begin{document} \begin{titlepage}
\rightline{\vbox{\halign{&#\hfil\cr
&SLAC-PUB-12580\cr
}}}
\begin{center}
\thispagestyle{empty} \flushbottom { {
\Large\bf Contact Interactions and Resonance-Like Physics at Present and Future Colliders from Unparticles
\footnote{Work supported in part
by the Department of Energy, Contract DE-AC02-76SF00515}
\footnote{e-mail:
rizzo@slac.stanford.edu}}}
\medskip
\end{center}

\centerline{Thomas G. Rizzo}
\vspace{8pt} 
\centerline{\it Stanford Linear
Accelerator Center, 2575 Sand Hill Rd., Menlo Park, CA, 94025}

\vspace*{0.3cm}

\begin{abstract}
High scale conformal physics can lead to unusual unparticle stuff at our low energies.  In this paper we discuss how 
the exchange of unparticles between Standard Model fields can lead to new contact interaction physics as well as a 
pseudoresonance-like structure, an unresonance, that might be observable at the Tevatron or LHC in the Drell-Yan channel. 
The specific signatures of this scenario are quite unique and can be used to easily identify this new physics given 
sufficient integrated luminosity. 
\end{abstract}



\renewcommand{\thefootnote}{\arabic{footnote}} \end{titlepage} 

%
%
%

In a recent paper, Georgi{\cite{Georgi:2007ek}} has speculated on the existence of a high scale conformal sector{\cite{Banks:1981nn}} 
which may couple to the various gauge and matter fields of the Standard Model(SM). This new sector can lead to important phenomenological 
consequences in our low energy world through the interactions of some new `stuff', termed unparticles, whose properties have begun to be 
explored in an ever increasing number of phenomenological analyses{\cite{Georgi:2007si,Cheung:2007ue,Luo:2007bq,Chen:2007vv,Ding:2007bm,Liao:2007bx,
Aliev:2007qw,Li:2007by,Duraisamy:2007aw,Lu:2007mx,Stephanov:2007ry,Fox:2007sy,Greiner:2007hr,Davoudiasl:2007jr,
Choudhury:2007js,Chen:2007qr,Aliev:2007gr,Mathews:2007hr,Zhou:2007zq,Ding:2007zw,Liao:2007ic,Chen:2007je,Feng,King}}. 
There are good reasons to believe that the physics of 
such unparticles may best be explored at TeV scale colliders. Signals at such machines can result either from unparticle emission in otherwise 
SM processes, which leads to a new source of missing energy, and/or their exchange between SM fields which leads to new contact-like 
interactions.  Extending earlier work{\cite {Georgi:2007si,Cheung:2007ue,Mathews:2007hr,Feng}}, in this paper we will explore how the exchange 
of unparticles between SM fields can be probed in Drell-Yan collisions at the Tevatron and LHC as well as at ILC. The well-understood Drell-Yan channel 
is a particularly clean one at hadron colliders and provides us with an almost background free laboratory to look for many kinds of new Terascale interactions. 
The breaking of the conformal symmetry near the electroweak scale through the SM Higgs vev{\cite {Fox:2007sy}} leads to a pseudoresonance-like 
structure at, \eg, the LHC, that we term an unresonance, which can be used to uniquely identify unparticles as source of the new physics. 

As discussed in the literature{\cite {Chen:2007qr}}, there are very many ways for unparticles to interact with SM fields and the particular choice 
of a subset of these to examine will depend on a number of assumptions. For example, the interaction of a spin-1 unparticle with a pair of ordinary, 
in principle different, fermions may be written as 
\begin{equation}
{1\over {\Lambda^{d-1}}}\bar f \gamma_\mu (c_LP_L+c_RP_R)\tilde f{\cal O}^\mu\,,
\end{equation}
where $\Lambda$ is an effective mass scale, $d$ is the non-canonical scaling dimension of the unparticle field, which is expected to lie in the range $1\leq d < 2$, 
and $c_{L,R}$ are assumed to be $O(1)$ coefficients; $P_{L,R}=(1\mp \gamma_5)/2$ are the helicity projection operators as usual.  
Recall that a free ordinary gauge field has $d=1$ leading to a dimensionless coupling. For a spin-0 unparticle on the otherhand, the corresponding helicity-preserving 
interaction takes the form{\cite{Georgi:2007ek}}   
\begin{equation}
{1\over {\Lambda^{d-2}}}\bar f \gamma_\mu (c_L'P_L+c_R'P_R)\tilde f\partial^\mu {\cal O}\,.
\end{equation}
Such an operator is relatively suppressed for two reasons: ($i$) due to the presence of the derivative an additional power of $\Lambda$ appears in the 
denominator and, more importantly, ($ii$) the  mass dimension compensation in the numerator is through 
a linear combination of the $f,\tilde f$ masses. Since the initial and/or final state fermions at both the hadron colliders and 
ILC are all quite light it will be essentially impossible to probe such SM-fermion unparticle scalar interactions at these machines. Thus we will only 
consider the spin-1 unparticle case in the discussion below. 

At the Tevatron and LHC, gluon initial states may also contribute in the Drell-Yan channel{\cite {Mathews:2007hr}} 
provided that they simultaneously couple to unparticles through an operator of the form 
\begin{equation}
\sim {1\over {\Lambda^{d+1}}}G^a_{\mu\alpha}G^\alpha_{a,\nu} {\cal O^{\mu\nu}}\,,
\end{equation}
where the `field strength' $ {\cal O^{\mu\nu}}=\partial^\mu {\cal O^{\nu}}-\partial^\nu {\cal O^{\mu}}$. However, in comparison to the coupling to 
fermion pairs, this interaction appears to be suppressed by two additional powers of $\Lambda$. Furthermore, this new interaction introduces 
another free parameter into the model describing the assumed relative strength of unparticle couplings to fermion and to gauge fields. We will briefly 
address these issues in the discussion below noting that for equal strength couplings this $gg$ contribution to Drell-Yan production is relatively suppressed in 
comparison to that arising from the $q\bar q$ initial state at both the Tevatron and LHC.

In what follows we will make some further assumptions about the nature of the unparticles and their interactions. Being SM singlets, the simplest possibility is 
to imagine that they couple {\it universally} in a flavor-blind manner to all of the SM fermion representations. In such a case all of the strong constraints 
that may arise from flavor-changing processes can be trivially avoided as they are naturally absent. This also implies that the couplings of the SM 
fermions to unparticles is purely vector-like and thus we can absorb the parameter combination $c_L+c_R$ above into the definition of $\Lambda$ for convenience. 
Furthermore, the unparticle coupling to the SM Higgs doublet, as discussed in Ref.{\cite {Fox:2007sy}}, leads to a breakdown of conformal invariance once the 
Higgs obtains a vev. In such a situation, the authors of Ref.{\cite {Fox:2007sy}} note that the propagator for the unparticle will be modified and in a simple 
toy model will behave as 
\begin{equation}
\sim {{A_d}\over {\sin d\pi}}~\theta(P^0)~\theta(P^2-\mu^2)~(P^2-\mu^2)^{d-2}~e^{i\pi (d-2)}\,,
\end{equation}
with $A_d$ being the numerical factor given by Georgi{\cite{Georgi:2007ek}}. We will assume that this form provides a reasonable description in this situation and 
we will discuss the implications of this form further below as it leads to striking collider signatures; we imagine, however, that the $\theta$ function threshold 
is somewhat smoothed out in a more realistic scenario but also that the turn-on of unparticle exchange remains very rapid. Note that for the processes of interest 
to us we do not need to assume that the unparticle is transverse, \ie, $\partial_\nu {\cal O^\nu}=0$, since the external fermions can be treated as massless to a 
very good approximation. 

\begin{figure}[htbp]
\centerline{
\includegraphics[width=7.5cm,angle=90]{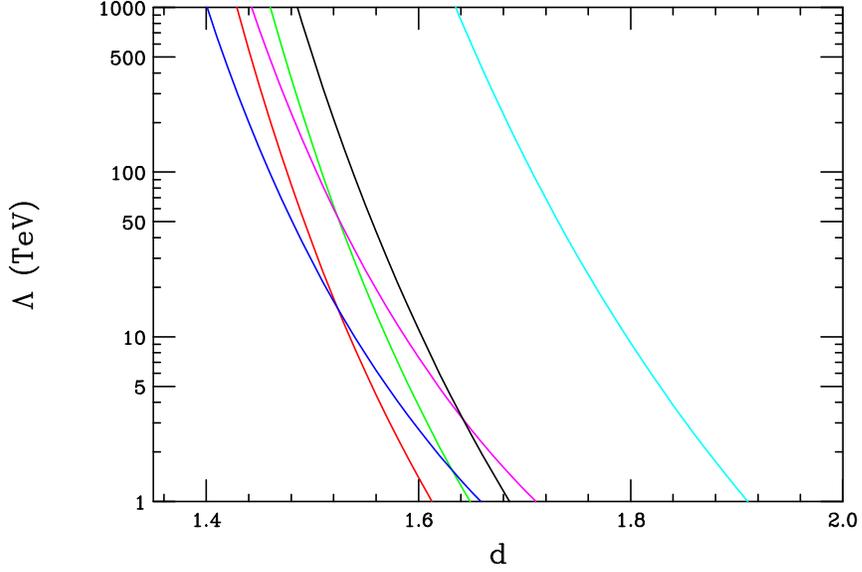}}
\vspace*{0.1cm}
\caption{Low energy bounds of the scale $\Lambda$ as a function of $d$ from $(g-2)_e$(red,green), $(g-2)_\mu$(blue,magenta) and orthopositronium decay(cyan,black). 
The first(second) member of each pair corresponds to purely vector(axial-vector) couplings. The excluded region is below and to the left of each curve.}
\label{fig1}
\end{figure}

In expression above $\mu$ is the effective `low' energy scale where conformal invariance is broken in the unparticle sector. Note that this 
expression implies that the presence of $\mu \neq 0$, even if $\mu$ is only a few GeV, cuts 
off the low-momentum unparticle exchange modes resulting in a very strong suppression of unparticle-induced contributions to many processes at low energies. 
These include, \eg, the contributions to $g-2$ for both $e$ and $\mu$, modifications to positronium decay as well as any of the remaining flavor-changing interactions 
involving $K,D$ and $B$ mesons. With such a  non-zero $\mu$, the important bounds on the scale $\Lambda$ arising from these types of processes can be completely 
evaded{\footnote {These bounds, when $\mu=0$, can be particularly strong as shown in Fig.~\ref{fig1}}}.  It is also possible that even those, as yet unexplored, 
constraints arising from precision electroweak measurements can also be evaded for reasonable values of $\mu$ while simultaneously allowing values of $\Lambda$ 
to be not too far above 1 TeV. In that case unparticle physics may remain accessible at Tevatron, LHC and ILC energies. 
Perhaps one might naturally expect that $\mu \sim 1$ TeV and so is not too distant in magnitude from that of the SM Higgs vev, $~246$ GeV, which breaks the 
conformal symmetry in the SM; we will assume that this possibility is realized within an approximate order of magnitude in the analysis presented below. As we will 
see the value of $\mu$ will play a very important role in collider searches for unparticle physics under the assumptions outlined above. 

Let us consider the contact-interaction-like contribution of unparticle exchange in the Drell-Yan process, $q\bar q \to \ell^+\ell^-$, as a function of the dilepton 
invariant mass, $M$. The parton-level triple differential cross section for this process can be easily calculated{\cite {Cheung:2007ue,Rizzo:2006nw}}. Note 
that in addition to the usual SM terms, this subprocess differential cross section picks up an extra unparticle induced piece proportional to the following 
combination of factors for each initial state quark (neglecting terms odd in the scattering angle here which integrate to zero in the total cross section):  
\begin{equation}
2fe^2Q_eQ_q \cos (d-2)\pi +2fv_e v_q M^2~{{(M^2-M_Z^2)\cos (d-2)\pi+M_Z\Gamma_Z\sin (d-2)\pi}\over {(M^2-M_Z^2)^2+(M_Z\Gamma_Z)^2}}+f^2\,,
\end{equation}
where $e^2=4\pi \alpha$, $Q_{e,q}$ are the electron and quark charges and $v_{e,q}$ are the vector couplings of the electron and quarks to the SM $Z$; 
with $\hat s=M^2$, we have defined   
\begin{equation}
f={{A_d}\over {\sin d\pi}}~\theta(\hat s-\mu^2) ~{{\hat s}\over {\Lambda^2}}~\Big[{{(\hat s-\mu^2)}\over {\Lambda^2}}\Big]^{d-2}\,,
\end{equation}
where $M_Z(\Gamma_Z)$ is the mass(width) of the SM $Z$. The factor $f$ arises directly from the assumed form of the unparticle propagator given above. 
Note that in the limit when $d=2$ we recover the kinematic structure of the ordinary, traditional dimension-6 contact interaction structure apart from a normalization 
factor{\cite{Eichten:1983hw}} and the 
$\theta$-function threshold. We now observe that for $\mu >0$ and $d<2$, if taken at face value, the factor $f$ leads to a very unusual threshold behavior 
for unparticle physics since $f$ vanishes when $\hat s <\mu^2$ but is very large when $\hat s$ is immediately above $\mu^2$. This means that a distorted peak will 
appear in the dilepton mass distribution in the region near $\hat s =\mu^2$ whose visibility will depend on the values of $d$ and $\Lambda$ as well as the detector 
mass resolution. In what follows we will consider only range $1<d<2$ since for $d\geq 2$ the propagator is not well defined. 

As discussed above the unparticle coupling to both gluons and fermions simultaneously leads to an additional contribution to the Drell-Yan process which must be 
added incoherently to that arising from the $q\bar q$ initial state. Using the same normalization as in that case, the parton level cross section for 
$gg \to \ell^+\ell^-$ is, apart from common overall factors, essentially given by
\begin{equation}
{3\over {32}}~c^2~f^2~\Big[{{\hat s}\over {{\Lambda}^2}}\Big]^2\,,
\end{equation}
where $c$ describes the relative strength of the $gg$ coupling to unparticles relative to that of the corresponding flavor universal 
$q\bar q$ coupling one. Note the extra suppression by the additional powers of $\Lambda$ 
which we anticipated in the discussion above. Numerically, from a detailed calculation we then find that for $c=1$ and $\Lambda \geq M \geq 1$ TeV at the LHC, this 
$gg$ contribution is only $\sim 5-10\%$ or less of that of the $q\bar q$ contribution to the total cross section; this relative contribution is even smaller 
at the Tevatron since in that case both the $q$ and $\bar q$ are valence quarks. These results are found to be independent of the other unparticle model 
parameters $d$ and $\mu$. Thus the $gg$ contribution can generally be neglected in the discussion that follows to a reasonably good approximation.

To match conditions at the Tevatron (LHC) as well as possible we employ approximate NNLO K-factors, apply rapidity cuts on the outgoing leptons, $|\eta_\ell|<1(2.5)$, 
and  Gaussian smearing to the dilepton mass distribution with a $\sim 1\%$ resolution to approximate that of, \eg, the CDF and ATLAS detectors for the electron 
final state{\cite {CDF,ATLAS}} in the relevant dilepton mass ranges. We can then compare the predictions of 
the SM with those of the unparticle scenario for different values of the parameters 
$d, \Lambda$ and $\mu$. The issue here is, of course, not just whether or not the effects of the unparticles are visible above the SM Drell-Yan background but whether 
the predictions for unparticle physics can be uniquely identified and also if the different values of the unparticle parameters are distinguishable at the Tevatron 
and LHC. Obtaining the value of $\mu$, given sufficient center of mass energy, will be the most straightforward of these measurements at any collider as it determines 
the unparticle threshold and the position of the corresponding resonance-like structure,

\begin{figure}[htbp]
\centerline{
\includegraphics[width=7.5cm,angle=90]{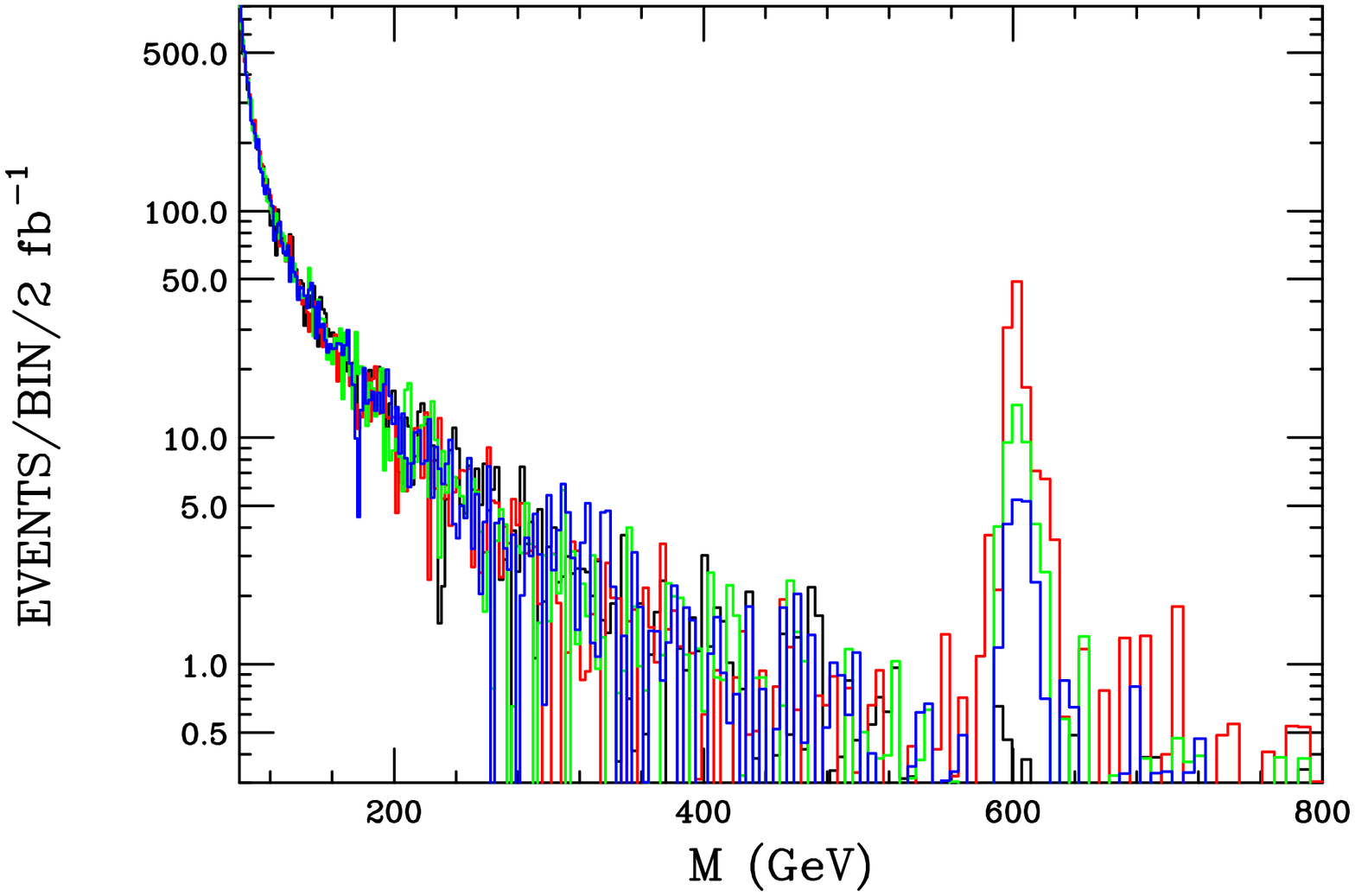}}
\vspace*{0.1cm}
\centerline{
\includegraphics[width=7.5cm,angle=90]{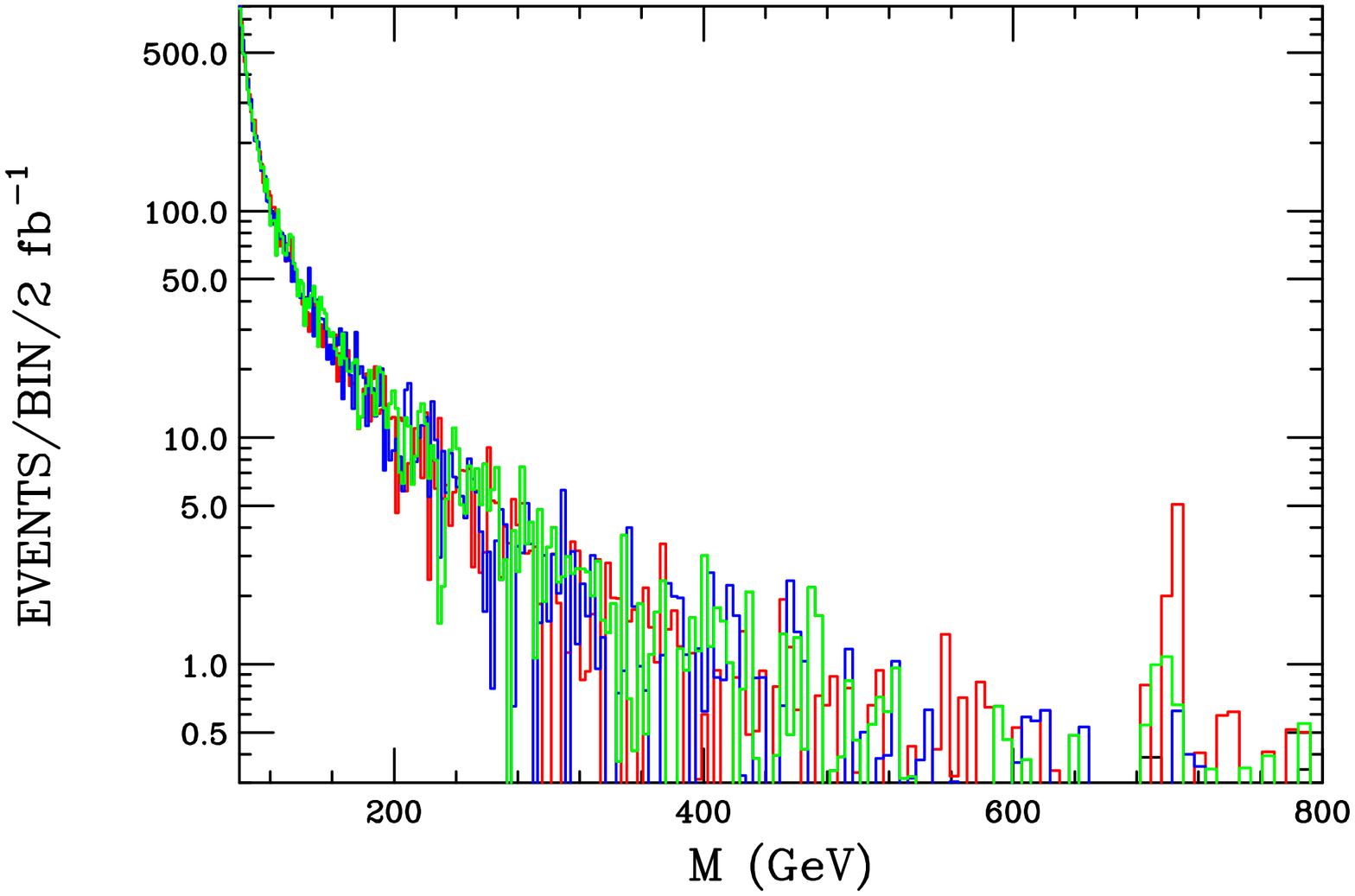}}
\vspace*{0.1cm}
\caption{Drell-Yan dilepton mass distribution at the Tevatron assuming $\mu=600(700)$ GeV with $d=1.5(1.6)$ in the top(bottom) panel. The red(green,blue) 
histogram corresponds to $\Lambda=1(2,3)$ TeV, respectively. The almost invisible black histogram is the SM prediction.}
\label{fig0}
\end{figure}
\begin{figure}[htbp]
\centerline{
\includegraphics[width=7.5cm,angle=90]{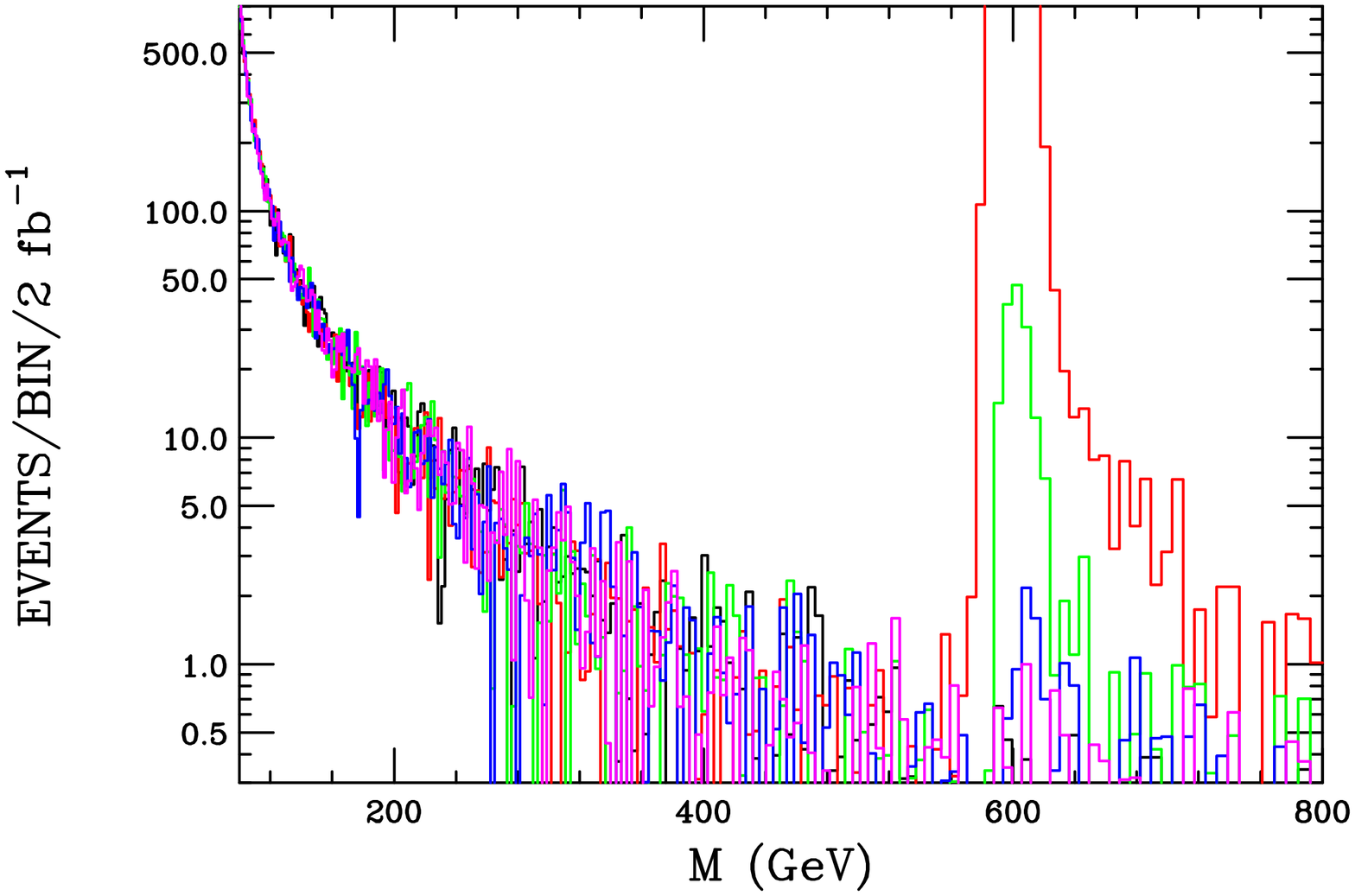}}
\vspace*{0.1cm}
\centerline{
\includegraphics[width=7.5cm,angle=90]{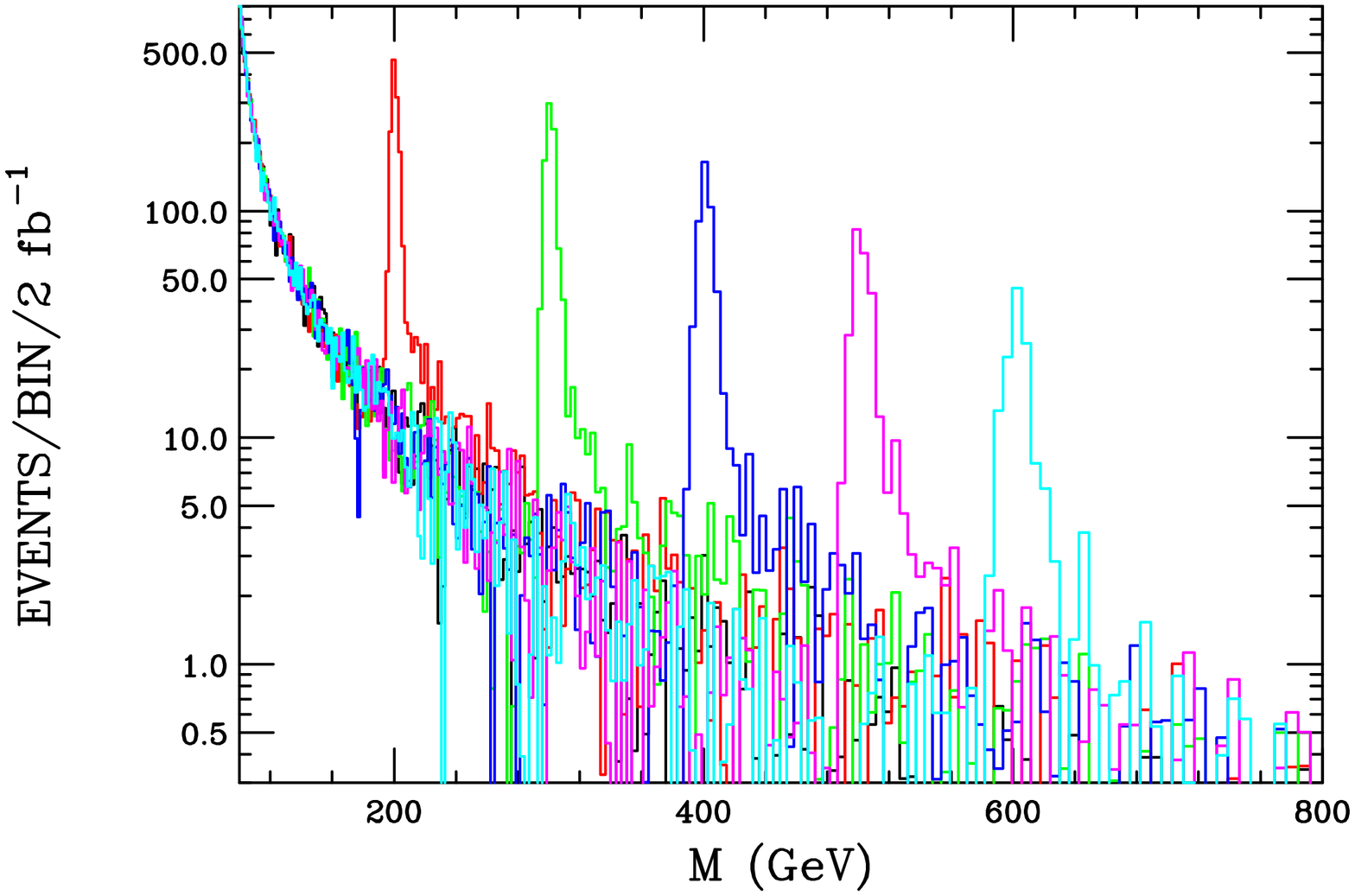}}
\vspace*{0.1cm}
\caption{(Top) Same as the previous figure but now with $\Lambda=1$ TeV and $\mu=600$ GeV for $d$=1.3(1.5,1.7,1.9) corresponding to the red(green,blue,magenta) 
histograms, respectively. (Bottom) In this case $\Lambda=1$ and $d=1.5$ with $\mu=200,300,400,500$ or 600 GeV. The SM prediction is the (almost invisible) 
black histogram in both panels.}
\label{fig00}
\end{figure}
\begin{figure}[htbp]
\centerline{
\includegraphics[width=7.5cm,angle=90]{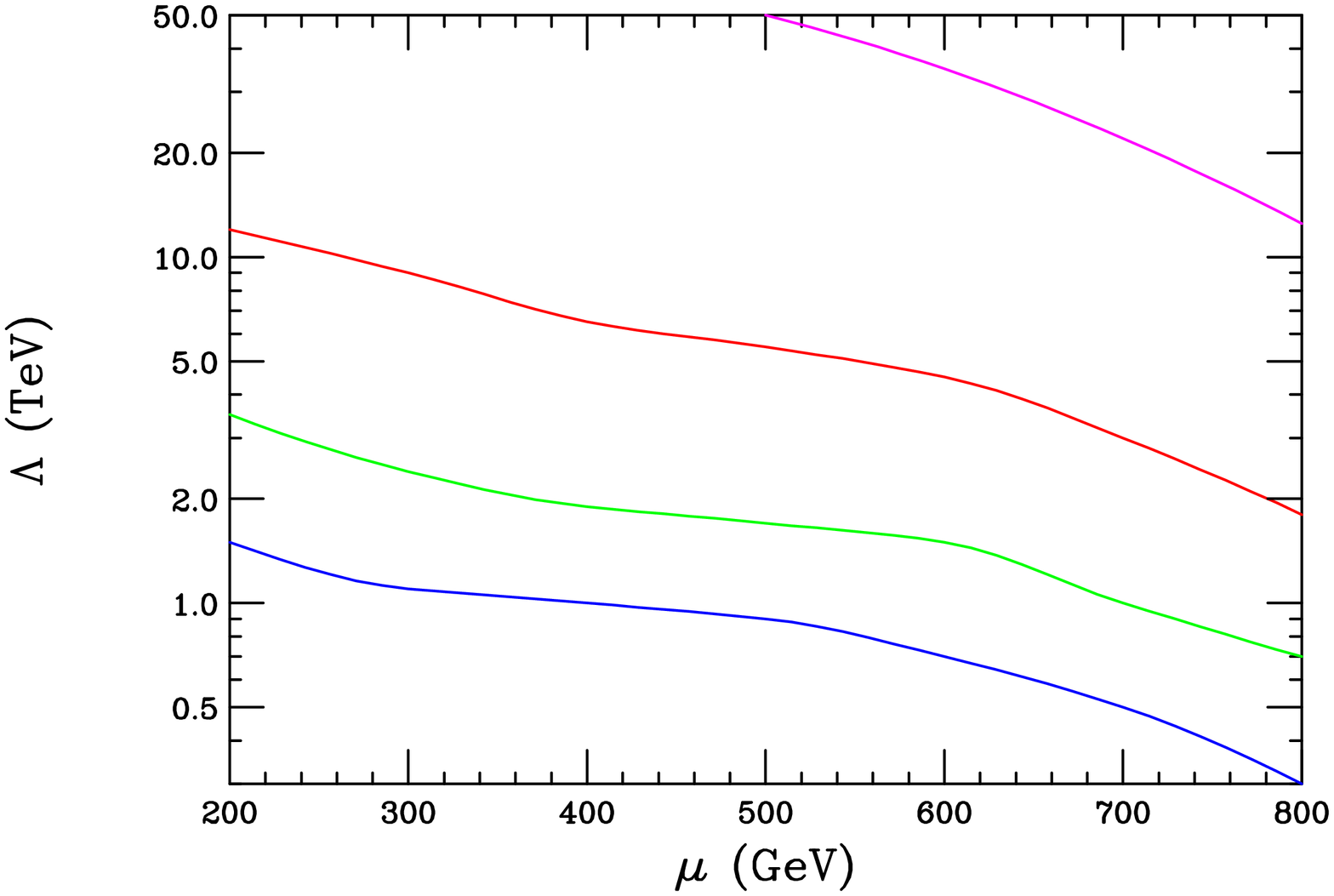}}
\vspace*{0.1cm}
\caption{Approximate lower bounds of the scale $\Lambda$ as a function of $\mu$ from the Tevatron. From top to bottom the curves correspond to $d=1.4$,1.5,1.6 and 
1.7, respectively}
\label{fig1p}
\end{figure}

Figs.~\ref{fig0} and ~\ref{fig00} show a selection of the predictions for this unparticle model at the Run II Tevatron. Here we see that the predictions of the SM are 
reproduced in all cases when $M < \mu$ but then a narrow peaking structure, the unresonance, is generated by the smeared 
singularity in the unparticle propagator. Note that there are 
{\it no} deviations from the SM until one sees the unresonance itself. This implies that indirect searches for this variety of unparticle physics are useless 
at values of $M$ below the resonance structure itself unlike, \eg, the indirect searches for more traditional resonances such as KK gravitons or new gauge 
bosons{\cite{Rizzo:2006nw}} which interfere with the SM in the sub-threshold mass regime. For example, at the ILC with a fixed center of mass energy  
no unparticle exchange signal would be observed unless $\mu < \sqrt s$ so that indirect constraints on the unparticle model parameters cannot be obtained 
using this quite standard technique. It is important to remember that although it 
generally appears as such, this unparticle structure is not an ordinary resonance like, \eg,  a $Z'$.  This unresonance is   
an artifact of the propagator threshold and large $M$ behavior combined with the finite detector resolution and the rapid fall off of the parton densities.

\begin{figure}[htbp]
\centerline{
\includegraphics[width=7.5cm,angle=90]{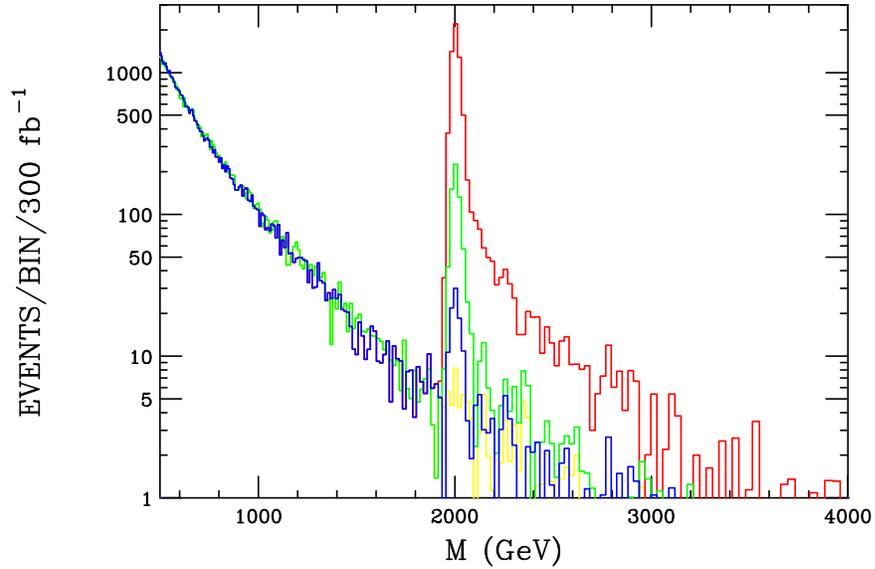}}
\vspace*{0.1cm}
\caption{High resolution scan of the Drell-Yan dilepton mass distribution at the LHC assuming a large integrated luminosity taking $d=1.5$, $\mu=2$ TeV with 
$\Lambda=3(10,25)$ TeV corresponding to the red(green,blue) histogram. The yellow histogram is the SM prediction. Adding the contribution from the $gg$ initial 
state makes an essentially invisible addition to this result.}
\label{fig000}
\end{figure}

It is clear from these figures that such a strange effect should be observable at the Tevatron for a reasonable range of the model parameters although the 
details of this structure 
are dependent on the specific choices made for the values of $d,\mu$ and $\Lambda$. The lack of observation of such a structure of any kind by the Tevatron  
clearly already constrains these model parameters. Certainly if $\mu$ is sufficiently small such an unresonance 
structure would have been already observed{\cite {limit}} unless $\Lambda$ was simultaneously also very large. For example, a simple analysis indicates 
that if $d=1.5$ then we must have $\mu>600$ GeV unless $\Lambda \gsim 4.5$ TeV. For smaller values of 
$\Lambda$, somewhat larger values of $\mu$ can be excluded.  Furthermore, these constraints are seen to become stronger(weaker) for smaller(larger) values of $d$. 
A rough summary of the possible Tevatron constraints on the model parameters from the lack of observation of any  
Drell-Yan resonances is shown in Fig.~\ref{fig1p}. It is interesting to note that if such an unresonance  
were to become visible at the Tevatron in the near future it would be somewhat difficult to distinguish it from a more ordinary resonance due to the low statistics 
available. It goes without saying that a detailed determination of the model parameters from such a resonance at the Tevatron would also be 
difficult if not impossible to obtain.  
As an aside, it is interesting to note that such an unresonance should also have appeared at LEP II in all final state fermion channels provided $\Lambda$ is 
not too large, $\lsim 5-10$ TeV, and this also suggests that $\mu \gsim 200$ GeV since the effective mass resolution of LEP detectors was somewhat better than those 
at the Tevatron. 

\begin{figure}[htbp]
\centerline{
\includegraphics[width=7.5cm,angle=90]{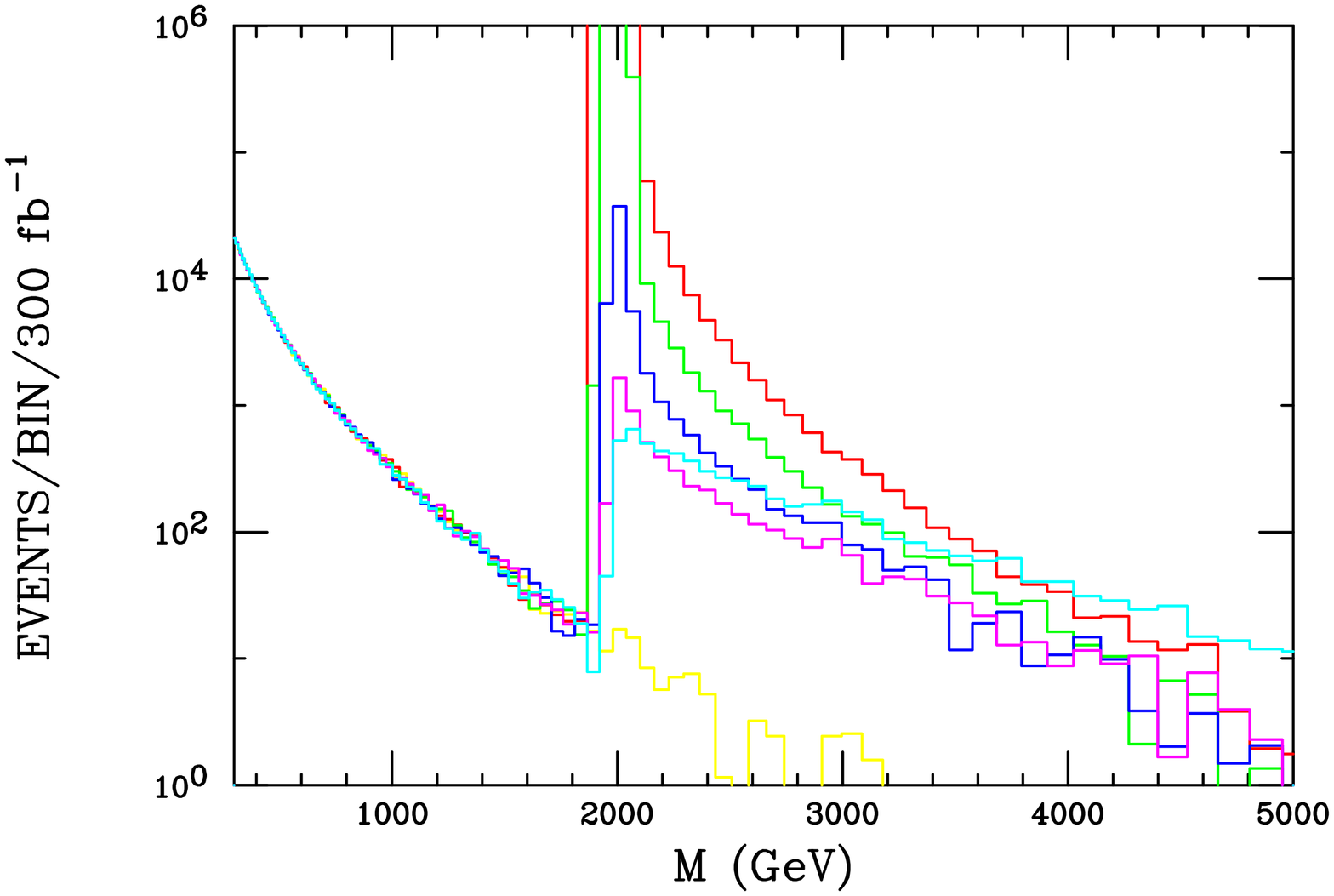}}
\vspace*{0.1cm}
\centerline{
\includegraphics[width=7.5cm,angle=90]{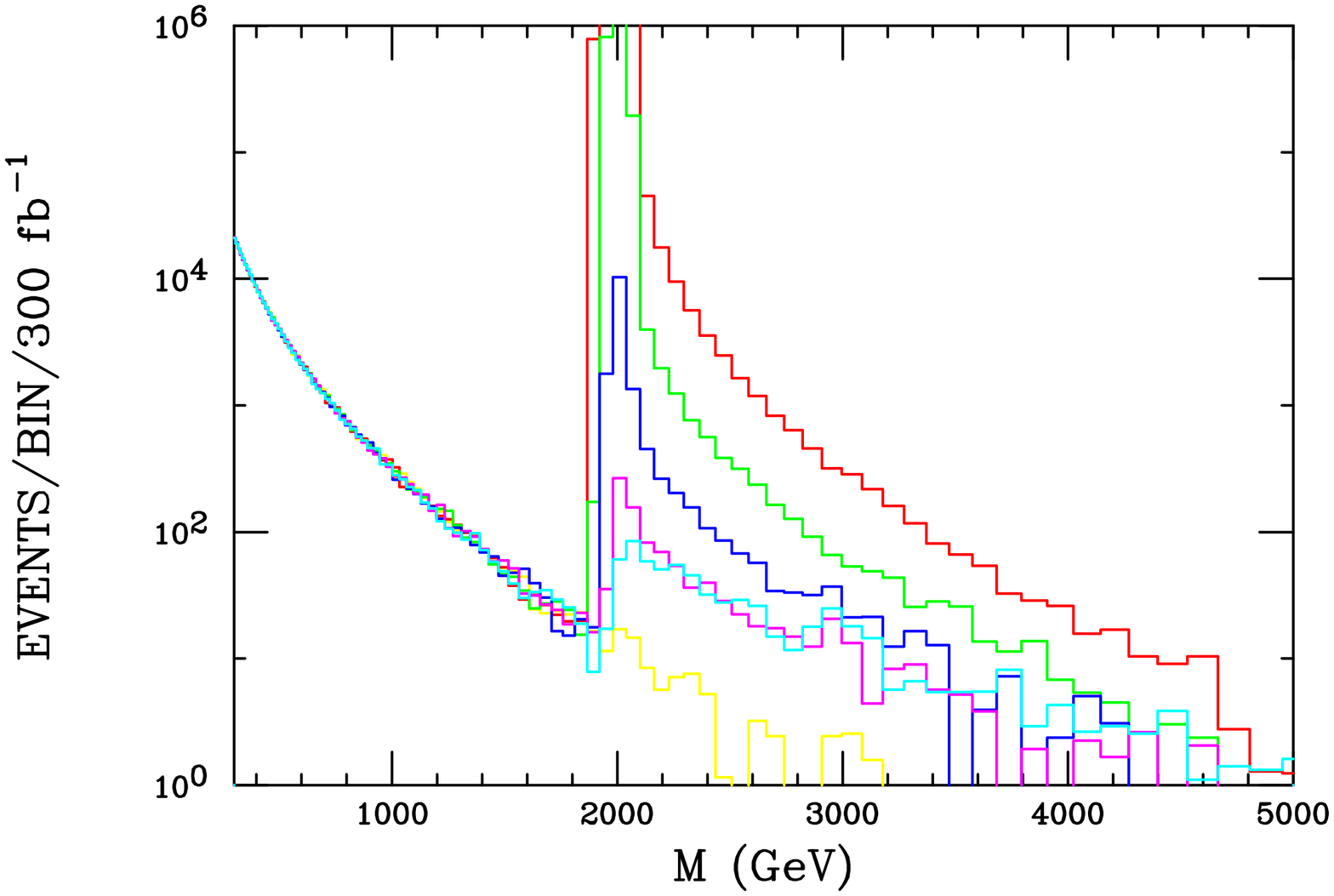}}
\vspace*{0.1cm}
\caption{Drell-Yan dilepton mass distribution at the LHC with high luminosity assuming $\mu=2$ TeV with $\Lambda=1(2)$ TeV in the top(bottom) panel. The yellow 
histogram is the SM prediction while the red(green,blue,magenta,cyan) histograms corresponds to $d$=1.1(1.3,1.5,1.7,1.9), respectively.}
\label{fig2}
\end{figure}

Let us now look ahead to the LHC where for simplicity we will restrict our attention to cases where $\mu \geq 1$ TeV.{\footnote {We note that smaller 
values of $\mu$ with significantly larger values of $\Lambda$ remain allowed by both the Tevatron and LEP II data discussed above.}} To get a first glimpse of what 
these unparticle thresholds 
may look like in detail we show in Fig.~\ref{fig000} a simple example assuming that $d=1.5$ and $\mu=2$ TeV with a large integrated luminosity 
and a fine-grained binning of the distribution comparable to the dilepton mass resolution. Unlike the Tevatron, where the unresonance peak does an excellent job at 
faking a more typical resonance structure, at the LHC 
we see that there is instead an abrupt cross section edge at $M\simeq \mu$ followed by a substantial tail at larger values of $M$. In this first sample result we find 
that such a bizarre structure will be observable even for $\Lambda$ values in excess of 25 TeV although the identification of its special nature is lost in this 
large $\Lambda$ limit due to low statistics. For more modest values of $\Lambda$ the cross section rise is still by more than an order of magnitude indicating that 
this unusual new physics will be difficult to miss in this particular channel.

\begin{figure}[htbp]
\centerline{
\includegraphics[width=7.5cm,angle=90]{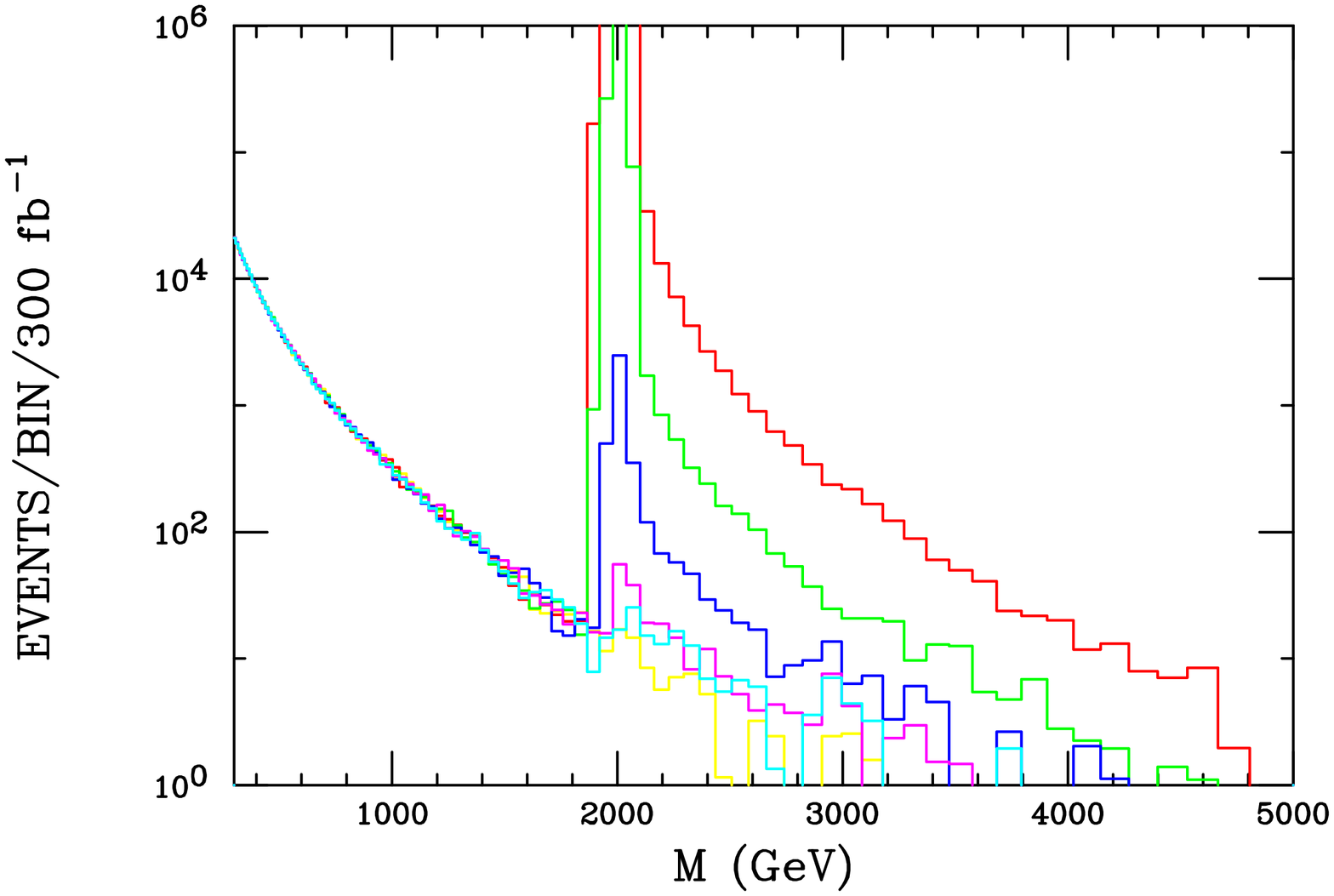}}
\vspace*{0.1cm}
\centerline{
\includegraphics[width=7.5cm,angle=90]{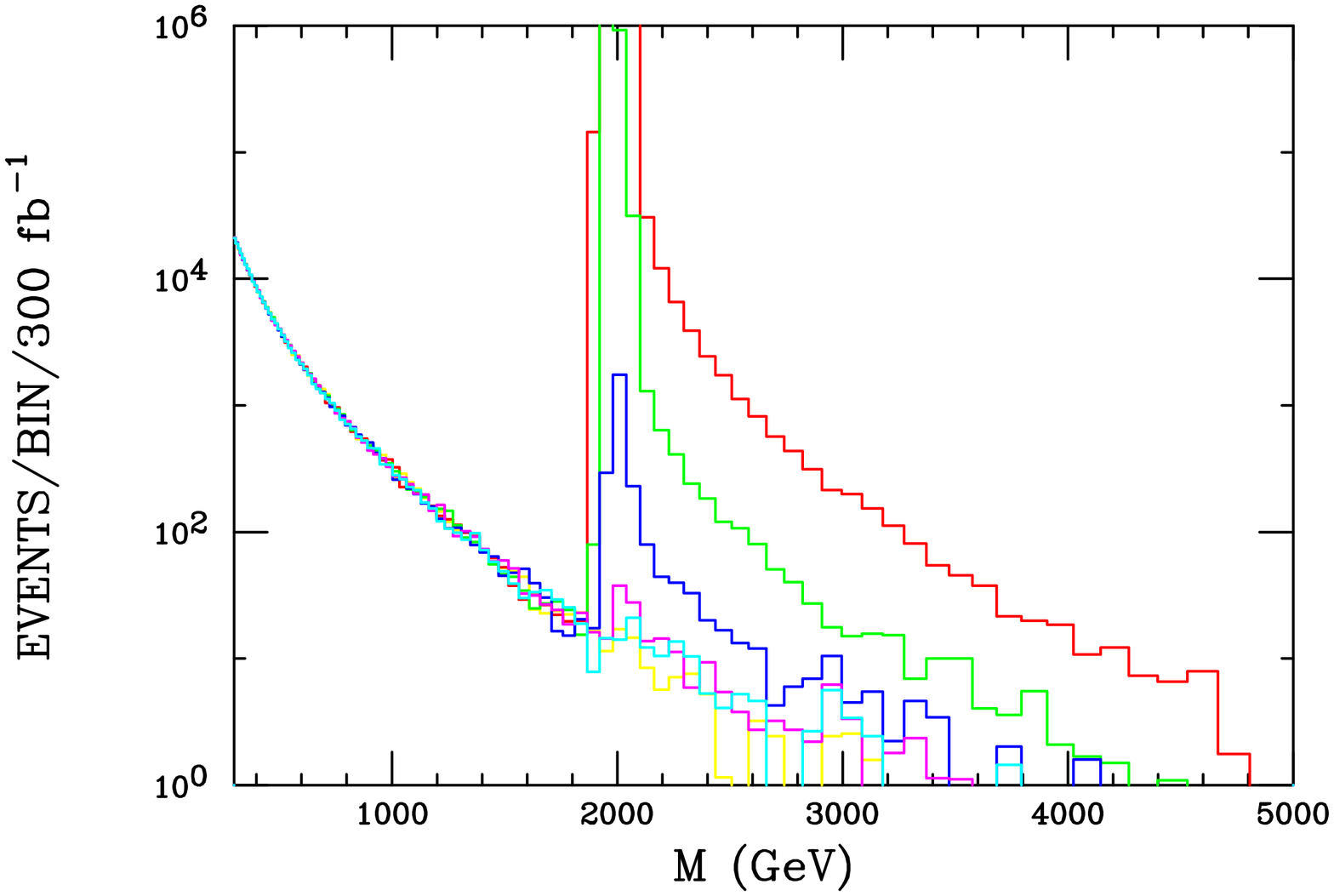}}
\vspace*{0.1cm}
\caption{Same as in the previous figure but now with $\Lambda=3(4)$ TeV in the top(lower) panel.}
\label{fig3}
\end{figure}

\begin{figure}[htbp]
\centerline{
\includegraphics[width=7.5cm,angle=90]{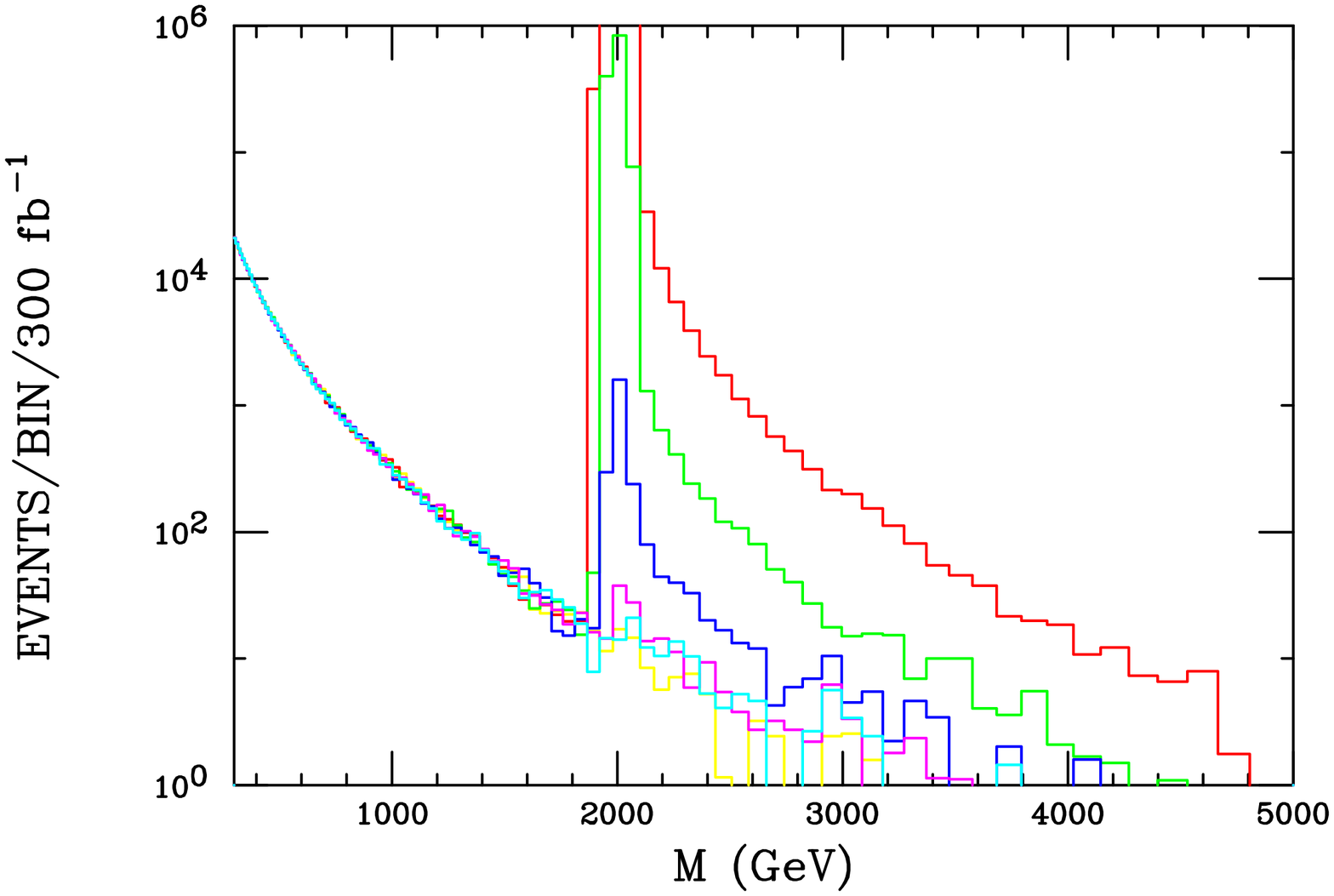}}
\vspace*{0.1cm}
\centerline{
\includegraphics[width=7.5cm,angle=90]{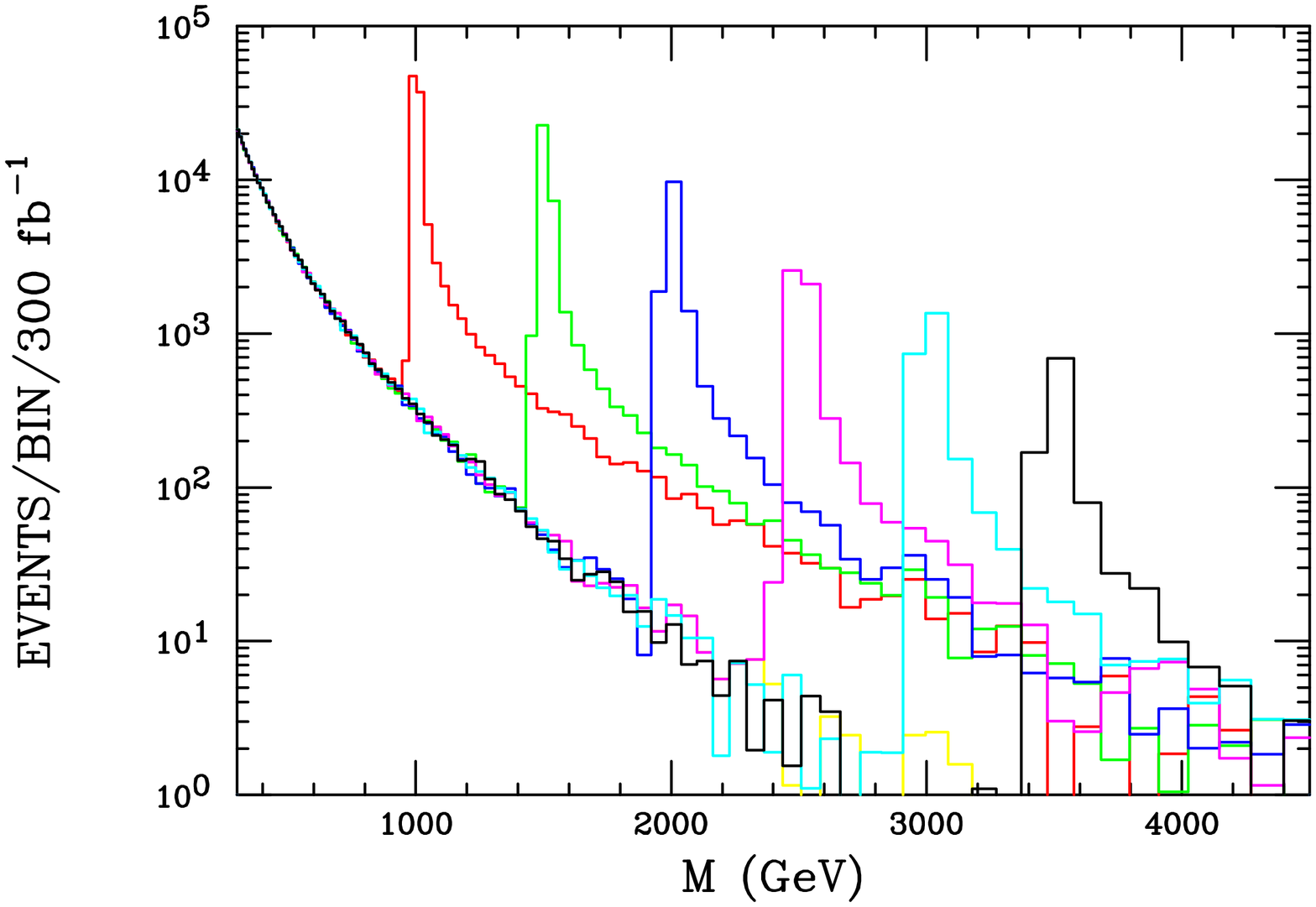}}
\vspace*{0.1cm}
\caption{In the top panel, same as the previous figure but now with $\Lambda=5$ TeV. In the lower panel, $d=1.5$ and $\Lambda=2$ TeV is 
assumed for the signal histograms which, from left to right, correspond to $\mu=1$,1.5,2,2.5,3 and 3.5 TeV, respectively.}
\label{fig4}
\end{figure}

\begin{figure}[htbp]
\centerline{
\includegraphics[width=7.5cm,angle=90]{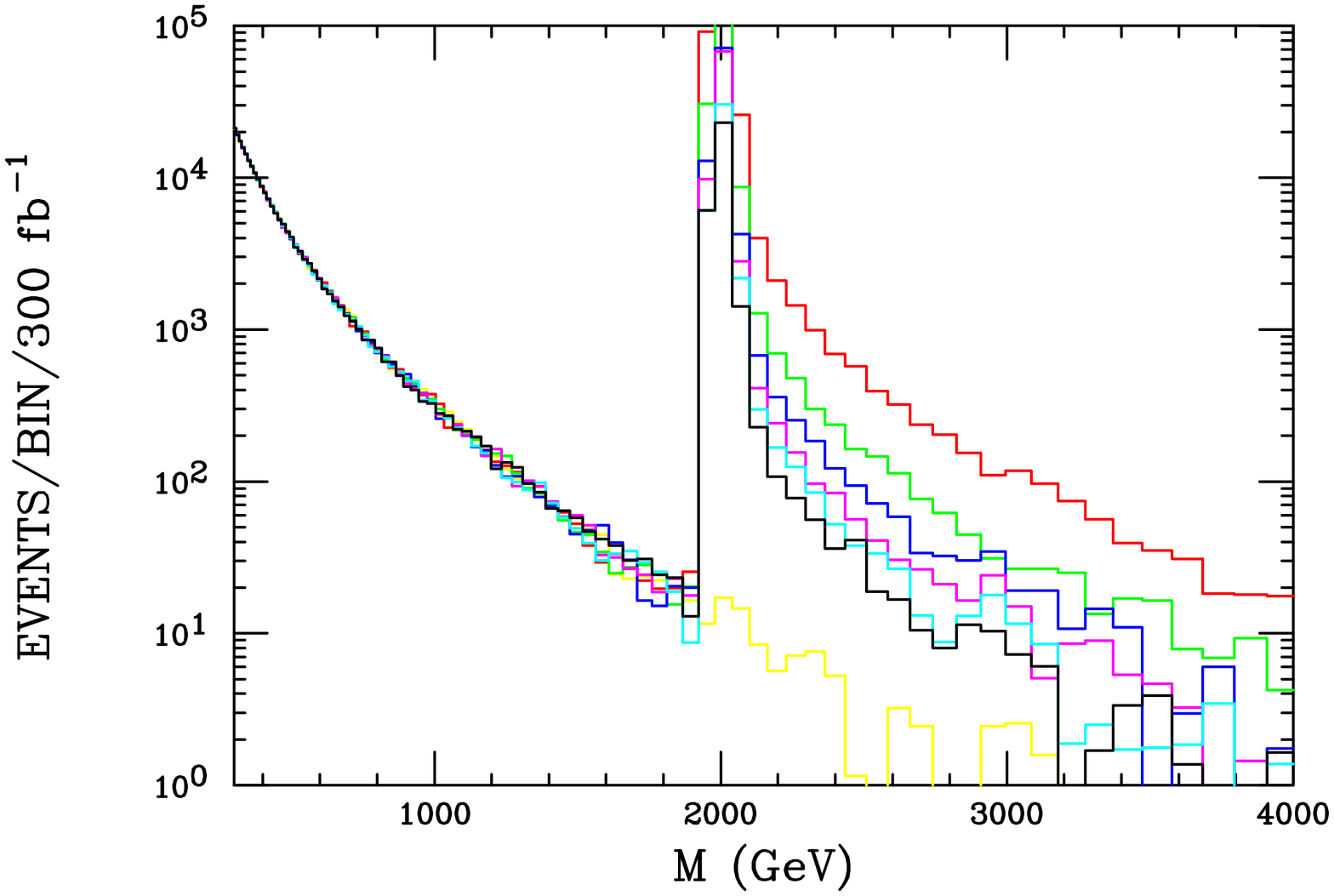}}
\vspace*{0.1cm}
\centerline{
\includegraphics[width=7.5cm,angle=90]{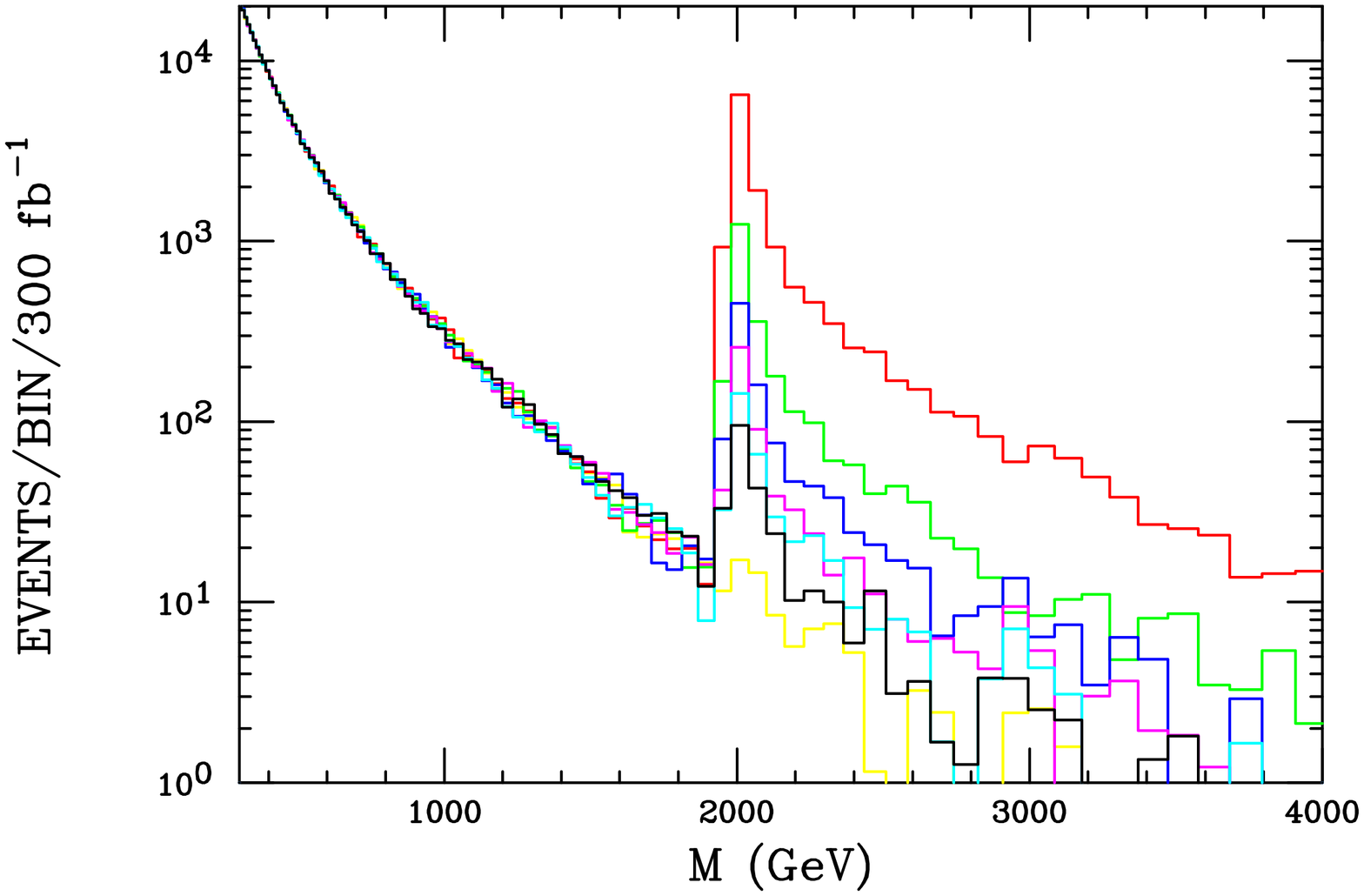}}
\vspace*{0.1cm}
\caption{Same as in Fig.~\ref{fig2} but now with $d$=1.4(1.6) and $\mu=2$ TeV in the top(bottom) panel. The signal histograms, from top to bottom, correspond to 
$\Lambda=1-6$ TeV in steps of 1 TeV.}
\label{fig5}
\end{figure}

Figs.~\ref{fig2},\ref{fig3},\ref{fig4} and \ref{fig5} show an array of unparticle model predictions for the LHC assuming a high integrated luminosity as the three  
parameters are varied independently. Several things 
are immediately clear from these results: ($i$) As expected, the abrupt increase in the cross section due to the unresonance allows for a rather precise determination 
of the value of $\mu$ over a very large range of the other parameters. This result, however, has relied on the use of the specific form of the threshold behavior 
described above. In a more realistic model where this threshold is smoother this determination may be somewhat more difficult if the appropriate scale is 
sufficiently high. However, for a low mass scale threshold, the statistics with 100 $fb^{-1}$ of integrated luminosity 
should be sufficiently large to precisely determine the details of this threshold 
structure. ($ii$) The specific details of the height of the peak and the shape of the tail 
of the unresonance structure itself is controlled by a combination of the both the $d$ and $\Lambda$ parameters. The height of the peak is found to be 
only weakly dependent on the value of $\Lambda$ itself when the corresponding value of $d$ is 
small, $\gsim 1$. However, this sensitivity grows rapidly as $d$ is increased and is found to scale as $\sim 1/\Lambda^{2(d-1)}$ for fixed $d$. On 
the otherhand, for fixed $\Lambda$, the unresonance peak height is expected to scale exponentially with $d$ which explains the strong $d$ dependence we observe 
in these figures. ($iii$) Note that the generic shape of the high energy tail of the distribution undergoes a significant qualitative change as the value of $d$ 
passes through $d=1.5$. Provided a signal of the unresonance is 
indeed observed at the LHC with sufficient statistics it would seem from these figures that it will be easily distinguished from a more traditional $Z'$-like 
resonance. Furthermore, it would also appear to be quite straightforward to extract the values of the parameters $\mu$, $d$ and $\Lambda$ with reasonable precision 
from the location and height of the peak itself as well as the shape of its high energy tail. Note that while the extraction of $\mu$ may be sensitive to the details 
of the modeling of the threshold region, the shape of the tail is not so that both $d$ and $\Lambda$ can be determined independently of the details of the threshold 
behavior. The details of how well this determination can be done would depend upon having a more realistic treatment of the threshold region.

\begin{figure}[htbp]
\centerline{
\includegraphics[width=7.5cm,angle=90]{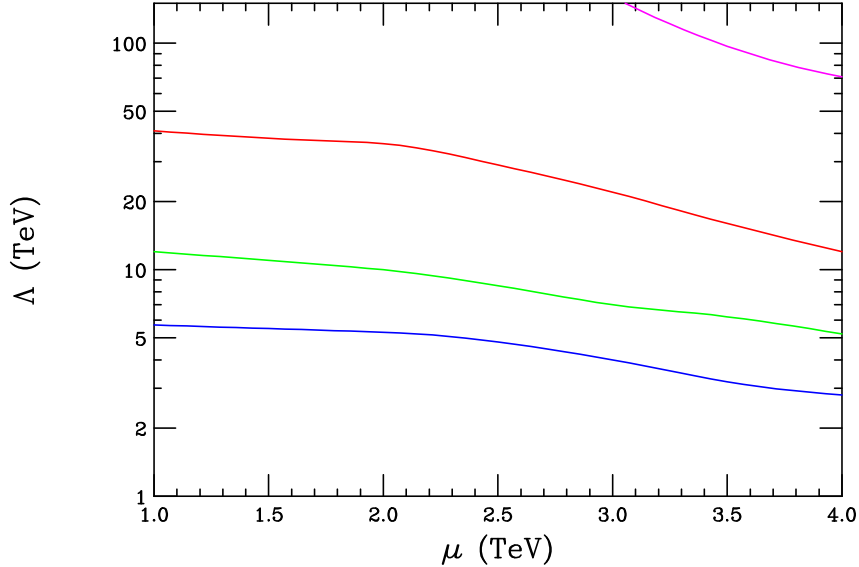}}
\vspace*{0.1cm}
\caption{$5\sigma$ search reach as in Fig.~\ref{fig1p} but now for the LHC assuming an integrated luminosity of 300 $fb^{-1}$.}
\label{fig6}
\end{figure}

One can also ask what ranges of the parameter space of the unparticle scenario can be explored at the LHC with high luminosity; the answer lies in Fig.~\ref{fig6}. 
Here we see that, \eg, with $\mu=2$ TeV and $d=1.5$, the LHC has significant sensitivity to unresonance production up to values of $\Lambda$ in the 30-40 TeV range. 
As expected this sensitivity is seen to decline(improve) as either $\mu$ or $d$ is increased(decreased). Of course, with a lower integrated luminosity the reach 
will not be as great but will still remain significantly better than that obtainable at the Tevatron.


In this paper we have explored the physics associated with spin-1 unparticle exchange in the Drell-Yan process, $gg,q\bar q \to \ell^+ \ell^-$, at both the Tevatron 
and the LHC. The new physics associated with these unparticles can be truly bizarre. The analysis presented here is based on two assumptions: ($i$) that the 
unparticle couples universally to all SM fermion representations and ($ii$) the propagator for 
the unparticle due to the Higgs-induced broken conformal symmetry was at least approximately that given in Ref.{\cite{Fox:2007sy}}. In this picture, all of the 
very strong low-energy constraints on unparticle exchanges are naturally avoided and the relevant scale $\Lambda$ may be as low as a few TeV and hence accessible to 
colliders. In our analysis we have focused on the well-defined parameter space 
region $1 \leq d \leq 2$. In this case the form of the unparticle propagator leads to a sharp threshold in the contribution of unparticles to all exchange processes 
including Drell-Yan. For $d$ in this range this structure appears pseudoresonance-like, whose location is determined by the propagator parameter $\mu$, which we 
term an unresonance. With low statistics the unresonance would appear to be a typical narrow resonance as one would imagine being produced by a weakly interacting $Z'$ 
state. However, the true nature of this structure was shown here to be quite distinct and detectable over a wide range of model parameters at both the Tevatron and the 
LHC given sufficient luminosity. Furthermore, the shape of the unresonance structure and its tail was shown to be sufficiently sensitive to the parameters $d$ and 
$\Lambda$ that their values could be determined from the data. 

Due to the, at least approximate, $\theta$-function turn on of the effects of unparticle exchanges, the new physics associated with such particles 
will not be detectable at energy scales below $\mu$; this is quite unlike the exchanges of more ordinary massive particles whose 
effects may be detected indirectly below threshold. This implies that the physics of unresonances will not be accessible at the ILC if $\mu >1$ TeV. 

It would be extremely interesting if such unphysics were to make its appearance at the LHC.

\noindent{\Large\bf Acknowledgments}

The author would like to thank JoAnne Hewett for discussions related to this paper.

%
\def\MPL #1 #2 #3 {Mod. Phys. Lett. {\bf#1},\ #2 (#3)}
\def\NPB #1 #2 #3 {Nucl. Phys. {\bf#1},\ #2 (#3)}
\def\PLB #1 #2 #3 {Phys. Lett. {\bf#1},\ #2 (#3)}
\def\PR #1 #2 #3 {Phys. Rep. {\bf#1},\ #2 (#3)}
\def\PRD #1 #2 #3 {Phys. Rev. {\bf#1},\ #2 (#3)}
\def\PRL #1 #2 #3 {Phys. Rev. Lett. {\bf#1},\ #2 (#3)}
\def\RMP #1 #2 #3 {Rev. Mod. Phys. {\bf#1},\ #2 (#3)}
\def\NIM #1 #2 #3 {Nuc. Inst. Meth. {\bf#1},\ #2 (#3)}
\def\ZPC #1 #2 #3 {Z. Phys. {\bf#1},\ #2 (#3)}
\def\EJPC #1 #2 #3 {E. Phys. J. {\bf#1},\ #2 (#3)}
\def\IJMP #1 #2 #3 {Int. J. Mod. Phys. {\bf#1},\ #2 (#3)}
\def\JHEP #1 #2 #3 {J. High En. Phys. {\bf#1},\ #2 (#3)}


\begin{thebibliography}{99}

\bibitem{Georgi:2007ek}
  H.~Georgi,
  arXiv:hep-ph/0703260.

\bibitem{Banks:1981nn}
  T.~Banks and A.~Zaks,
  Nucl.\ Phys.\  B {\bf 196}, 189 (1982).

\bibitem{Georgi:2007si}
  H.~Georgi,
  arXiv:0704.2457 [hep-ph].

\bibitem{Cheung:2007ue}
  K.~Cheung, W.~Y.~Keung and T.~C.~Yuan,
  arXiv:0704.2588 [hep-ph].

\bibitem{Luo:2007bq}
  M.~Luo and G.~Zhu,
  arXiv:0704.3532 [hep-ph].

\bibitem{Chen:2007vv}
  C.~H.~Chen and C.~Q.~Geng,
  arXiv:0705.0689 [hep-ph].

\bibitem{Ding:2007bm}
  G.~J.~Ding and M.~L.~Yan,
  arXiv:0705.0794 [hep-ph].

\bibitem{Liao:2007bx}
  Y.~Liao,
  arXiv:0705.0837 [hep-ph].

\bibitem{Aliev:2007qw}
  T.~M.~Aliev, A.~S.~Cornell and N.~Gaur,
  arXiv:0705.1326 [hep-ph].

\bibitem{Li:2007by}
  X.~Q.~Li and Z.~T.~Wei,
  arXiv:0705.1821 [hep-ph].

\bibitem{Duraisamy:2007aw}
  M.~Duraisamy,
  arXiv:0705.2622 [hep-ph].

\bibitem{Lu:2007mx}
  C.~D.~Lu, W.~Wang and Y.~M.~Wang,
  arXiv:0705.2909 [hep-ph].

\bibitem{Stephanov:2007ry}
  M.~A.~Stephanov,
  arXiv:0705.3049 [hep-ph].

\bibitem{Fox:2007sy}
  P.~J.~Fox, A.~Rajaraman and Y.~Shirman,
  arXiv:0705.3092 [hep-ph].

\bibitem{Greiner:2007hr}
  N.~Greiner,
  arXiv:0705.3518 [hep-ph].

\bibitem{Davoudiasl:2007jr}
  H.~Davoudiasl,
  arXiv:0705.3636 [hep-ph].

\bibitem{Choudhury:2007js}
  D.~Choudhury, D.~K.~Ghosh and Mamta,
  arXiv:0705.3637 [hep-ph].

\bibitem{Chen:2007qr}
  S.~L.~Chen and X.~G.~He,
  arXiv:0705.3946 [hep-ph].

\bibitem{Aliev:2007gr}
  T.~M.~Aliev, A.~S.~Cornell and N.~Gaur,
  arXiv:0705.4542 [hep-ph].

\bibitem{Mathews:2007hr}
  P.~Mathews and V.~Ravindran,
  arXiv:0705.4599 [hep-ph].

\bibitem{Zhou:2007zq}
  S.~Zhou,
  arXiv:0706.0302 [hep-ph].

\bibitem{Ding:2007zw}
  G.~J.~Ding and M.~L.~Yan,
  arXiv:0706.0325 [hep-ph].

\bibitem{Chen:2007je}
  C.~H.~Chen and C.~Q.~Geng,
  arXiv:0706.0850 [hep-ph].

\bibitem{Liao:2007ic}
  Y.~Liao and J.~Y.~Liu,
  arXiv:0706.1284 [hep-ph].

\bibitem{Feng}
  M.~Bander, J.L.~Feng, A.~Rajaraman, and Y.~Shirman,
  arXiv:0706.2677 [hep-ph].

\bibitem{King}
  K.~Cheung, W.-Y.~Keung and T.-C.~Yuan,
  arXiv:0706.3155 [hep-ph].


\bibitem{Rizzo:2006nw}
  For some background and basic formalism, see, T.~G.~Rizzo,
  arXiv:hep-ph/0610104.

\bibitem{CDF}
  T.~Kamon,
  M.~G.~Albrow {\it et al.}  [CDF Collaboration],
  Nucl.\ Instrum.\ Meth.\  A {\bf 480}, 524 (2002).


\bibitem{ATLAS}
CMS Physics TDR, Volume II: CERN-LHCC-2006-021,
ATLAS TDR, http://atlas.web.cern.ch/Atlas/GROUPS/PHYSICS/TDR/TDR.html.
 
\bibitem{Eichten:1983hw}
  E.~Eichten, K.~D.~Lane and M.~E.~Peskin,
  Phys.\ Rev.\ Lett.\  {\bf 50}, 811 (1983).

\bibitem{limit}
The direct search lower limit on , \eg, the mass of  a new $Z'$ gauge boson with SM
couplings is approaching 1 TeV from Run II data at the Tevatron. The lower bound on a SM-coupled $Z'$ is 923 GeV from CDF.
See, for example,
P.~Savard, ``Searches for Extra Dimensions and New Gauge Bosons at the Tevatron,''
talk given at the {\it XXXIII International Conference on High Energy Physics},
26 July-2 August 2006, Moscow, Russia;
T.~Adams, ``Searches for New Phenomena with Lepton Final States at the Tevatron,'' 
talk given at {Rencontres de Moriond
Electroweak Interactions and Unified Theories 2007}, La Thuile, Italy 10-17 March 2007.




\end{thebibliography}
\end{document}